\documentclass[twocolumn,aps,pra,showpacs,superscriptaddress]{revtex4}

\usepackage{amsmath}
\usepackage{amssymb}

\begin{document}

\title{Decay process of quantum open system at finite-temperature }

\author{XIAO Xiao}

\affiliation{School of Applied Sciences, Beijing University of Technology, Beijing,
100124, China}

\author{GAO Yi-Bo}

\email{ybgao@bjut.edu.cn}

\affiliation{School of Applied Sciences, Beijing University of Technology, Beijing,
100124, China}

\begin{abstract}
Starting from the formal solution to the Heisenberg equation, we revisit
an universal model for a quantum open system with a harmonic oscillator
linearly coupled to a boson bath. The analysis of the decay process
for a Fock state and a coherent state demonstrate that this method
is very useful in dealing with the problems in decay process of the
open system. For finite temperature, the calculations of the reduced
density matrix and the mean excitation number for the open system
show that an initial coherent state will evolve into a temperature-dependant
coherent state after tracing over the bath variables. Also in short-time
limit, a temperature-dependant effective Hamiltonian for the open
system characterizes the decay process of the open system.
\end{abstract}


\maketitle

\section{Introduction}

Actually, no quantum system can be isolated from its surrounding environment
completely~\cite{Zurek83,Breuer02}. When it is coupled to the environment,
the system (also named open system) energy dissipates into the environment
irreversibly. There exists two approaches to deal with the decay process
of the open system, i.e., the system plus bath model~\cite{Leggett83}
and effective Hamiltonian model~\cite{Kanai48,Caldirola41}. In a
simple model of a harmonic oscillator interacting with a bath of many
harmonic oscillators, the relation between these two approaches is
clarified by Yu and Sun~\cite{Sun94,Sun95}.

In quantum optics and macroscopic quantum phenomena, the mesoscopic
or macroscopic states can be represented by coherent states. Due to
the quantum fluctuation and thermal fluctuation of the bath, it is
very important to study how the open system evolves with an initial
Fock state~\cite{Ninglu89} and an initial coherent state~\cite{Walls85},
especially at finite temperature.

Usually, the master equation method is adopted to treat the decay
process of the open system~\cite{Carmichael91}, time evolution of
the reduced matrix for the open system can be solved analytically
and numerically. In this paper, starting from the formal solution
to the Heisenberg equation, we can construct the reduced density matrix
for the open system, this method has been applied
to deal with some problems in quantum open system~\cite{Sun94,Sun95,Liu00,Sun98,Gao05,Gao07}.
Here, an exactly solvable model of one boson interacting with a bath
of many bosons in rotating-wave approximation is used to analyze the
decay process of the open system.

The paper is organized as follows. In sec.II, we describe an universal
model for quantum open system, a single mode boson interacting with
a bath of many bosons through system-bath coupling. In sec.III, we
revisit the decay process of a Fock state and a coherent state at
zero temperature. In sec.IV, we study the finite-temperature decay
process of the open system. Finally, we summarize our results.

\section{system-bath model}

The quantum decay process is characterized with the vanishing of diagonal
elements of the time evolution of reduced density matrix for an open
system. The simplest model is adopted that the composite system consists
of the system (a single mode boson) interacting with the bath of multi-mode
bosons. Using the method in Refs.~\cite{Sun94,Sun95,Liu00,Sun98,Gao05,Gao07}, we can construct the reduced
density matrix and calculate the mean excitation number for the open system.

The model Hamiltonian for this composite system is
\begin{equation}
H=\hbar\omega_{b}b^{\dag}b+\sum_{j}\hbar\omega_{j}a_{j}^{\dag}a_{j}+\sum_{j}\xi_{j}\left(ba_{j}^{\dag}+b^{\dag}a_{j}\right),\label{eq:Ham-t}
\end{equation}
where $b$ ($a_{j}$) and $b^{\dagger}$ ($a_{j}^{\dagger}$) are
bosonic annihilation and creation operators for the system (bath),
and the $\xi_{j}$ are complex coupling constants parameterized via
the spectral density function $J\left(\omega\right)$, i.e.,
\begin{equation}
J\left(\omega\right)=\sum_{j}\left|\xi_{j}\right|^{2}\delta\left(\omega-\omega_{j}\right).\label{eq:spectra function}
\end{equation}
Here we have applied rotating-wave approximation on the model Hamiltonian
$H$.

Now we can use the time evolution of operator
in Heisenberg picture to construct a wavefunction (or a reduced density
matrix). These well-known solutions of Heisenberg operators are given
in Ref.\cite{Sun98}
\begin{eqnarray*}
b\left(t\right) & = & u\left(t\right)b\left(0\right)+\sum_{j}v_{j}\left(t\right)a_{j}\left(0\right),\\
a_{j}\left(t\right) & = & e^{-i\omega_{j}t}a_{j}\left(0\right)+u_{j}\left(t\right)b\left(0\right)+\sum_{s}v_{j,s}\left(t\right)a_{s}\left(0\right).
\end{eqnarray*}
Where some coefficients in the above equation are
\begin{eqnarray}
u\left(t\right) & = & e^{-\frac{\gamma t}{2}}e^{-i\omega_{b}t},\nonumber \\
v_{j}\left(t\right) & = & \xi_{j}e^{-i\omega_{j}t}\frac{e^{-\frac{\gamma t}{2}}e^{-i\left(\omega_{b}-\omega_{j}\right)t}-1}{\omega_{b}-\omega_{j}-i\frac{\gamma}{2}},\nonumber \\
u_{j}\left(t\right) & = & \xi_{j}^{\ast}e^{-i\omega_{j}t}\frac{e^{-\frac{\gamma t}{2}}e^{-i\left(\omega_{b}-\omega_{j}\right)t}-1}{\omega_{b}-\omega_{j}-i\frac{\gamma}{2}}.\label{eq:coefficients-H}
\end{eqnarray}
When the spectral function of the bath is assumed as
\begin{equation}
J\left(\omega\right)=\frac{\gamma}{\pi},\label{eq:spectral-dissipation}
\end{equation}
the calculations show that
\begin{equation}
\sum_{j}\left\vert v_{j}\left(t\right)\right\vert ^{2}=1-e^{-\gamma t}.\label{dissipation relation}
\end{equation}
and $\left\vert u\left(t\right)\right\vert ^{2}+\sum_{j}\left\vert v_{j}\left(t\right)\right\vert ^{2}=1$
which maintains the commutation relation $\left[b\left(t\right),b^{\dagger}\left(t\right)\right]=1$.

\section{zero-temperature decay process}

In quantum optics, the study of Fock state and coherent state is very
important. At zero temperature, the decay of the Fock state and coherent
state is presented with master equation method in Refs.~\cite{Ninglu89,Walls85}.
Here, using the method in Refs.~\cite{Sun94,Sun95,Liu00,Sun98,Gao05,Gao07}, we revisit the decay of a Fock state and a coherent state.

\subsection{Decay of a Fock state}

The initial state of the composite system is assumed in
\begin{equation}
\left\vert \psi\left(0\right)\right\rangle =\left\vert n\right\rangle \otimes\left\vert \left\{ 0_{j}\right\} \right\rangle ,\label{initial state}
\end{equation}
where $n$ stands for the excitation number for the Fock state $\left\vert n\right\rangle $
and the bath is in vacuum state $\left\vert \left\{ 0_{j}\right\} \right\rangle $.
The ground state of the composite system is $\left\vert 0\right\rangle \otimes\left\vert \left\{ 0_{j}\right\} \right\rangle $,
which is invariant under the operation of Unitary evolution operator
$U\left(t\right)$, i.e., $U\left(t\right)\left\vert 0\right\rangle \otimes\left\vert \left\{ 0_{j}\right\} \right\rangle =\left\vert 0\right\rangle \otimes\left\vert \left\{ 0_{j}\right\} \right\rangle $.
Then we can construct the time evolution of the composite system,
\begin{eqnarray*}
\left\vert \psi\left(t\right)\right\rangle  & = & U\left(t\right)\left\vert \psi\left(0\right)\right\rangle \\
 & = & \frac{\left(B^{\dag}\left(t\right)\right)^{n}}{\sqrt{n!}}\left\vert 0\right\rangle \otimes\left\vert \left\{ 0_{j}\right\} \right\rangle .
\end{eqnarray*}
Where the operator $B\left(t\right)$ is defined as $B\left(t\right)=U\left(t\right)b\left(0\right)U^{\dag}\left(t\right)$.
Using the method in Ref.\cite{Gao05}, we can replace $i$ with $-i$
in Eq.(\ref{eq:coefficients-H}) and get
\begin{equation}
B^{\dag}\left(t\right)=u\left(t\right)b^{\dag}\left(0\right)+\sum_{j}v_{j}\left(t\right)a_{j}^{\dag}\left(0\right).\label{eq:anti-Heisenberg operator}
\end{equation}

Tracing over the bath variables in density matrix for the composite
system $\rho\left(t\right)=\left\vert \psi\left(t\right)\right\rangle \left\langle \psi\left(t\right)\right\vert $,
the reduced density matrix for the open system is calculated as
\[
\rho_{s}\left(t\right)=\sum_{m=0}^{n}P_{m}\left(t\right)\left\vert m\right\rangle \left\langle m\right\vert
\]
where the probability in the state $\left|m\right\rangle $ for $m=0...n$
is
\begin{equation}
P_{m}\left(t\right)=\frac{n!}{\left(n-m\right)!m!}e^{-m\gamma t}\left(1-e^{-\gamma t}\right)^{n-m}\label{eq:prob-Fock}
\end{equation}
Here we have taked into account the relation
\begin{equation}
\left\langle \left\{ 0_{j}\right\} \right\vert \left(\sum_{i}A_{i}\right)^{k}\left(\sum_{j}A_{j}^{\dagger}\right)^{l}\left\vert \left\{ 0_{j}\right\} \right\rangle =k!\left(1-e^{-\gamma t}\right)^{k}\delta_{kl},\label{eq:delta-relation}
\end{equation}
where $A_{j}=v_{j}^{*}\left(t\right)a_{j}$$\left(0\right)$. According
to the result in Eq.(\ref{eq:prob-Fock}), we obtain the probability
in the state $\left\vert n\right\rangle $, $P_{n}\left(t\right)=e^{-n\gamma t}$
(seen in Ref.~\cite{Ninglu89}). Here the decay time of the open
system is defines by
\begin{equation}
\tau^{-1}=n\gamma.\label{eq:decay time-Fock}
\end{equation}

\subsection{Decay of a coherent state}

Similar as the decay of a Fock state, for an initial coherent state
$\left\vert \alpha\right\rangle $, we construct the time evolution
of wavefunction for the composite system,

\begin{eqnarray*}
\left\vert \psi\left(t\right)\right\rangle  & = & e^{\alpha B^{\dag}\left(t\right)-\alpha^{\ast}B\left(t\right)}\left\vert 0\right\rangle \otimes\left\vert \left\{ 0_{j}\right\} \right\rangle \\
 & = & \left\vert \alpha u\left(t\right)\right\rangle \otimes\prod\limits _{j}\left\vert \alpha v_{j}\left(t\right)\right\rangle .
\end{eqnarray*}
Then the corresponding reduced density matrix is
\[
\rho_{s}\left(t\right)=\left\vert \alpha u\left(t\right)\right\rangle \left\langle \alpha u\left(t\right)\right\vert .
\]
Through the calculation of the mean excitation number for the open
system,
\[
\overline{N}=Tr\left(\rho_{s}\left(t\right)b^{\dag}b\right)=\left|\alpha\right|^{2}e^{-\gamma t},
\]
the decay time of a coherent state $\left|\alpha\right\rangle $ is
obtained (seen in Ref.\cite{Walls85}),
\begin{equation}
\tau^{-1}=\gamma.\label{eq:decay time-coherent}
\end{equation}

When the bath is in excited state (coherent state $\left\vert \left\{ \lambda_{j}\right\} \right\rangle $),
the master equation method is not valid. Using our method, we can easily obtain the time evolution of wavefunction for
the composite,
\begin{eqnarray}
\left\vert \psi\left(t\right)\right\rangle  & = & e^{\alpha B^{\dag}-\alpha^{\ast}B}\left\vert 0\right\rangle \otimes\prod\limits _{j}e^{\lambda_{j}A_{j}^{\dag}-\lambda_{j}^{\ast}A_{j}}\left\vert 0_{j}\right\rangle \nonumber \\
 & = & \left\vert \mu\left(t\right)\right\rangle \otimes\left\vert \left\{ \mu_{j}\left(t\right)\right\} \right\rangle .\label{eq:coherent-state bath}
\end{eqnarray}
where the two coefficients in above equation are
\begin{eqnarray*}
\mu\left(t\right) & = & \alpha u\left(t\right)+\sum_{j}\lambda_{j}u_{j}\left(t\right),\\
\mu_{j}\left(t\right) & = & \alpha v_{j}\left(t\right)+\lambda_{j}e^{-i\omega_{j}t}+\sum_{s\neq j}v_{s,j}\left(t\right).
\end{eqnarray*}

\section{finite-temperature decay process}

In previous section, it shows that when the bath is prepared in a
vacuum state, the quantum vacuum fluctuation of the bath will induce
decay process of the open system. Here we will demonstrate the finite-temperature
influence on decay process of the open system when the bath is in
a thermal equilibrium state.

Initially, the density operator for the composite system is written
as a direct product
\[
\rho\left(0\right)=\left(\left\vert \psi\left(0\right)\right\rangle \left\langle \psi\left(0\right)\right\vert \right)\otimes\rho_{B}.
\]
Here the initial state of the open system is assumed in a coherent
state, $\left\vert \psi\left(0\right)\right\rangle =\left\vert \alpha\right\rangle $.
In coherent-state representation, the bath at thermal equilibrium
is described by the density operator
\[
\rho_{B}=\prod\limits _{j}\int\frac{d^{2}\lambda_{j}}{\pi\overline{n_{j}}}e^{-\left\vert \lambda_{j}\right\vert ^{2}/\overline{n_{j}}}\left\vert \lambda_{j}\right\rangle \left\langle \lambda_{j}\right\vert ,
\]
where the mean excitation number in the $j$-th mode of the bath with
the frequency $\omega_{j}$ is
\begin{equation}
\overline{n_{j}}=\left(e^{\beta\hbar\omega_{j}}-1\right)^{-1}.\label{eq:thermal mean number}
\end{equation}

Using the result in Eq.(\ref{eq:coherent-state bath}),
the reduced density matrix for the open system is calculated,
\begin{equation}
\rho_{s}\left(t\right)=\prod\limits _{j}\left[\int\frac{d^{2}\lambda_{j}}{\pi\overline{n_{j}}}e^{-\left\vert \lambda_{j}\right\vert ^{2}/\overline{n_{j}}}\left\vert \mu\left(t\right)\right\rangle \left\langle \mu\left(t\right)\right\vert \right].\label{eq:reduced density-T}
\end{equation}
It can be used to characterize the decay process of the open system.
Then the probability of the open system in the initial state $\left|\psi\left(0\right)\right\rangle =\left|\alpha\right\rangle $,
i.e., diagonal elements of the reduced density matrix, can be obtained
\[
\left\langle \psi\left(0\right)\right\vert \rho_{s}\left(t\right)\left\vert \psi\left(0\right)\right\rangle =\left\langle \alpha\right|\left(\left\vert \psi_{s}\left(t\right)\right\rangle \left\langle \psi_{s}\left(t\right)\right\vert \right)\left|\alpha\right\rangle .
\]
Here the temperature-dependant wavefunction for the open system is
obtained,
\begin{equation}
\left\vert \psi_{s}\left(t\right)\right\rangle =\Phi\left(T,t\right)^{-\frac{1}{2}}\left\vert \alpha\left(\left(u\left(t\right)-1\right)\Phi\left(T,t\right)^{-\frac{1}{2}}+1\right)\right\rangle .\label{wavefunction-T-1}
\end{equation}
It shows that at finite temperature the state of the system evolves
into a coherent state $\left\vert \psi_{s}\left(t\right)\right\rangle $
according to a coherent state ($\left|\alpha\right\rangle $) initially.
In Eq.(\ref{wavefunction-T-1}), the time-(temperature-)dependant
term is denoted as
\begin{equation}
\Phi\left(T,t\right)=1+\sum_{j}\overline{n_{j}}\left\vert u_{j}\left(t\right)\right\vert ^{2}.\label{n-j}
\end{equation}
Changing the sum into an integral in the above equation, we have
\[
\sum_{j}\overline{n_{j}}\left\vert u_{j}\left(t\right)\right\vert ^{2}=\frac{\gamma}{2\pi}\int d\omega\frac{\left|e^{-\frac{\gamma t}{2}-i\left(\omega_{b}-\omega\right)t}-1\right|^{2}}{\left(\omega_{b}-\omega\right)^{2}+\left(\frac{\gamma}{2}\right)^{2}}\left(e^{\beta\hbar\omega}-1\right)^{-1}.
\]
Here the term $\left(e^{\beta\omega}-1\right)^{-1}$ is a slow-varying
function of the frequency $\omega$, thus we can take it out of the
integral and get
\[
\Phi\left(T,t\right)=1+\overline{n_{th}}\left(1-e^{-\gamma t}\right),
\]
where the mean excitation number with the frequency $\omega_{b}$
is $\overline{n_{th}}=\left(e^{\beta\hbar\omega_{b}}-1\right)^{-1}$.

To demonstrate the decay process of the open system, we now calculate
the mean excitation number for the open system. A decay process means
that the mean excitation number reduces with the time increasing.
Applying the time evolution of the wavefunction for the open system
in Eq.(\ref{wavefunction-T-1}), we can calculate the mean excitation
number to characterize the decay process of the open system, i.e.,
\begin{eqnarray*}
\overline{N} & = & \left\langle \psi_{s}\left(t\right)\right\vert b^{\dagger}b\left\vert \psi_{s}\left(t\right)\right\rangle \\
 & = & \left\vert \alpha\right\vert ^{2}\left\vert \left(u\left(t\right)-1\right)\Phi\left(T,t\right)^{-1}+\Phi\left(T,t\right)^{-\frac{1}{2}}\right\vert ^{2}.
\end{eqnarray*}

For low temperature ($\beta\rightarrow\infty$), we approximately
have $\overline{n_{th}}\simeq e^{-\beta\hbar\omega_{b}}$. Then the
mean excitation number of the open system becomes $\overline{N}\simeq\left\vert \alpha\right\vert ^{2}e^{-\gamma t}.$
It shows that, in the low-temperature limit, the above mean excitation
number $\overline{N}$ exponentially decays as the time $t$ increases
and approximately does not depend the temperature $T$. That is to
say, the vacuum fluctuations dominate the decay process of the open
system.

For high temperature ($\beta\rightarrow0$), we have $\overline{n_{th}}\simeq\left(\beta\hbar\omega_{b}\right)^{-1}$
and then the excitation number becomes
\[
\overline{N}\simeq\beta\hbar\omega_{b}\left\vert \alpha\right\vert ^{2}\left(1-e^{-\gamma t}\right)^{-1}\propto\frac{1}{T}.
\]
It shows that, in the high-temperature limit, the rise of the temperature
will accelerate the decay of the open system.

Assume the system initially in a state $\left\vert \psi_{s}\left(0\right)\right\rangle $,
it can be expanded into the superposition of a series of coherent
states, i.e.,
\begin{equation}
\left\vert \psi_{s}\left(0\right)\right\rangle =\int\frac{d^{2}\alpha}{\pi}C_{\alpha}\left\vert \alpha\right\rangle ,\label{any initial state}
\end{equation}
where the amplitude of the probability in the state $\left|\alpha\right\rangle $
is $C_{\alpha}=\left\langle \alpha|\psi_{s}\left(0\right)\right\rangle $.
Then the corresponding reduced density matrix will be
\begin{equation}
\rho_{s}\left(t\right)=\int\frac{d^{2}\alpha}{\pi}C_{\alpha}\left\vert \psi_{s,\alpha}\left(t\right)\right\rangle \int\frac{d^{2}\beta}{\pi}C_{\beta}^{\ast}\left\langle \psi_{s,\beta}\left(t\right)\right\vert ,\label{any-state}
\end{equation}
where for $\xi=\alpha,\beta$, we have defined
\[
\left\vert \psi_{s,\xi}\left(t\right)\right\rangle =\Phi\left(T,t\right)^{-\frac{1}{2}}\left\vert \xi\left(\left(u\left(t\right)-1\right)\Phi\left(T,t\right)^{-\frac{1}{2}}+1\right)\right\rangle .
\]
In short-time limit, $\gamma t\ll1$ and $\Phi\left(T,t\right)\simeq e^{\overline{n_{th}}\gamma t}$,
the temperature-dependant wavefunction of the open system becomes
\begin{equation}
\left\vert \psi_{s,\xi}\left(t\right)\right\rangle =e^{-iH_{eff}t}\left\vert \xi\right\rangle .\label{wavefunction-T-2}
\end{equation}
According to the results in Eq.(\ref{any-state},\ref{wavefunction-T-2}),
the corresponding temperature-dependant reduced density matrix for
the open system in Eq.(\ref{any-state}) can be rewritten into
\[
\rho_{s}\left(t\right)=e^{-iH_{eff}t}\left\vert \psi_{s}\left(0\right)\right\rangle \left\langle \psi_{s}\left(0\right)\right\vert e^{iH_{eff}t},
\]
where the temperature-dependant effective Hamiltonian for the open
system is obtained,
\begin{equation}
H_{eff}=\left(\omega_{b}-i\frac{\gamma}{2}\right)b^{\dag}b-i\frac{1}{2}\overline{n_{th}}\gamma.\label{eq:Ham-eff-T}
\end{equation}
Consider the the open system initially in a Fock state, $\left\vert \psi_{s}\left(0\right)\right\rangle =\left\vert n\right\rangle $,
the corresponding time evolution will be
\[
\left\vert \psi_{s}\left(t\right)\right\rangle =e^{-iH_{eff}t}\left\vert n\right\rangle =e^{-in\omega_{b}t}e^{-\frac{1}{2}\left(\overline{n_{th}}+n\right)\gamma t}\left\vert n\right\rangle
\]
and the mean excitation number becomes
\begin{equation}
\overline{N}\simeq ne^{-\left(\overline{n_{th}}+n\right)\gamma t}\label{eq:mean number-T-Fock}
\end{equation}
in which the decay time of the open system is
\begin{equation}
\tau^{-1}=\left(\overline{n_{th}}+n\right)\gamma.\label{eq:decay time-T-Fock}
\end{equation}

In a same fashion, consider the open system initially in a coherent
state, $\left\vert \psi_{s}\left(0\right)\right\rangle =\left\vert \alpha\right\rangle $,
the open system will evolve into a temperature-dependant coherent
state
\[
\left\vert \psi_{\alpha}\left(t\right)\right\rangle =e^{-iH_{eff}t}\left\vert \alpha\right\rangle =e^{-\frac{1}{2}\overline{n_{th}}\gamma t}\left\vert \alpha e^{-i\left(\omega_{b}-i\frac{\gamma}{2}\right)t}\right\rangle .
\]
Then the corresponding mean excitation number is calculated as
\begin{equation}
\overline{N}=\left\vert \alpha\right\vert ^{2}e^{-\left(\overline{n_{th}}+1\right)\gamma t}\label{mean number-T-coherent}
\end{equation}
with the decay time
\begin{equation}
\tau^{-1}=\left(\overline{n_{th}}+1\right)\gamma.\label{decay time-T-coherent}
\end{equation}
The results in Eq.(\ref{eq:decay time-T-Fock}) and Eq.(\ref{decay time-T-coherent})
show that thermal mean excitation number of the bath $\overline{n_{th}}$
accelerates the decay of the open system. When at zero temperature,
$\overline{n_{th}}=0$, the results in Eq.(\ref{eq:decay time-T-Fock})
and Eq.(\ref{decay time-T-coherent}) come back to the results at
zero temperature in Eq.(\ref{eq:decay time-Fock}) and Eq.(\ref{eq:decay time-coherent}).

\section{conclusions}

In this paper, we have studied an universal model for a quantum open
system with system-bath coupling. Using the method in Refs.~\cite{Sun94,Sun95,Liu00,Sun98,Gao05,Gao07},
not master equation method, we revisit the decay process of the open
system at zero temperature. Specially in short-time limit, a temperature-dependant
effective Hamiltonian for the open system is obtained (seen in Eq.(\ref{eq:Ham-eff-T})).
For finite temperature, it can characterize the decay process of the
open system very well. The decay time shows that the temperature $T$
will accelerate the decay process of the open system. In addition,
a temperature-dependant wavefunction for the open system is easily
obtained (seen in Eq.(\ref{wavefunction-T-1})), it shows that at
finite temperature, the open system will evolve into a temperature-dependant
coherent state from an initial coherent state. In summary, this method in Refs.~\cite{Sun94,Sun95,Liu00,Sun98,Gao05,Gao07} is very useful to deal with the problems of quantum
open system.

\begin{acknowledgments}
We acknowledge the support of the NSFC (Grant No.10604002). Y. B.
Gao would also like thank Prof. Yu-xi Liu for this discussions.
\end{acknowledgments}

\end{document}